\documentstyle[preprint,aps,psfig]{revtex}

\def\bea{\begin{eqnarray}}
\def\eea{\end{eqnarray}}
\def\be{\begin{eqnarray}}
\def\ee{\end{eqnarray}}
\def\ave#1{{\langle #1 \rangle}}

\begin{document}
\title{Event-by-event fluctuations of the charged particle ratio from
non-equilibrium transport theory}

\author{M.~Bleicher${}^{\xi}$, 
S. Jeon,
V. Koch
}

\address{Nuclear Science Division,\\
        Lawrence Berkeley National Laboratory,\\
        Berkeley, CA 94720, USA}

\footnotetext{${}^\xi$ Feodor Lynen Fellow of the Alexander v. Humboldt
  Foundation}{
\footnotetext{E-mail: bleicher@nta2lbl.gov, SJeon@lbl.gov, vkoch@lbl.gov}


\maketitle
\begin{abstract}
The event by event fluctuations of the ratio of positively to negatively
charged hadrons are predicted within the UrQMD model.
Corrections for finite acceptance and finite net charge are derived. These
corrections are relevant to compare experimental data and transport model
results to previous predictions.  
The calculated fluctuations at RHIC and SPS energies are shown to be 
compatible with a hadron gas. Thus, deviating by a factor of 3
from the predictions for a thermalized quark-gluon plasma.
\end{abstract}

\newpage

In this note we exploit a novel idea of a signal proposed by
\cite{jeon-koch} based on event by event fluctuations of the $N_+/N_-$ ratio
($N_+$ and $N_-$ being positively and negatively charged hadrons).
In particular, we would like to precisely specify the observable which
is to be compared with the prediction of Ref.\cite{jeon-koch}.
The magnitude of these fluctuations was estimated from thermodynamics
to be a factor of 3 different for a QGP compared  to a conventional 
hadronic gas. The predictions of \cite{jeon-koch} were
derived using a grand canonical ensemble and assuming a zero net charge. 
In a grand canonical ensemble charge is conserved only on the average, whereas
charge is conserved event-by-event in real heavy ion collisions and the
model calculations presented here.
The assumption of a grand canonical ensemble thus is only valid in the limit
of $P \equiv \ave{N_{\rm ch}}_{\Delta y} / \ave{N_{\rm ch}}_{\rm total}
\rightarrow 0$
and correction to that are of the type $(1 - P)$ 
\cite{jeon-koch,heiselberg-jackson}. Thus, in order to compare experimental
data with the predictions of \cite{jeon-koch}, corrections have to be applied,
which will be presented here.

For our study of this observable, we employ the Ultra-relativistic Quantum
Molecular Dynamics model (UrQMD) \cite{urqmd}. This model has been shown to
reproduce the measured data on event-by-event fluctuations in Pb+Pb collisions
at the SPS \cite{me}.
UrQMD is a microscopic transport approach based on the 
covariant propagation of constituent quarks and diquarks 
accompanied by mesonic and baryonic degrees of freedom.
It simulates multiple interactions of ingoing and newly produced 
particles, the excitation and fragmentation of color strings 
and the formation and decay of hadronic resonances. 
At SPS and RHIC energies, the treatment of subhadronic degrees of freedom is
of major importance. In the UrQMD model, these degrees of freedom enter via
the introduction of a formation time for hadrons produced in the 
fragmentation of strings \cite{andersson87a}.
The leading hadrons of the fragmenting strings contain the valence-quarks 
of the original excited hadron. In UrQMD they are allowed to
interact even during their formation time, with a reduced cross section
defined by the additive quark model,  thus accounting for the 
original valence quarks contained in that hadron \cite{urqmd}. 
The thermodynamics properties of the model in equilibrium are similar to a
Hagedorn gas with a limiting temperature of 135~MeV \cite{belkacem}.

Let us shortly review the main ideas of the charge fluctuations as
elaborated in \cite{jeon-koch}.
The following connection between the observable charge ratio 
fluctuations and the fluctuations of the elementary net 
charges of the system has been proposed: 
\be
D =
\langle N_{\rm ch}\rangle\langle \delta R^2\rangle 
\approx 4 \langle \delta Q^2\rangle / \langle N_{\rm ch}\rangle\quad,
\label{eq:dR2eqdQ2}
\ee
with $Q=\sum_a q_a N_a$, where $a$ runs over all hadron/parton species in the
interval $\Delta y \ll y_{\rm total}$ 
in an event and $q_a$ is the charge of the 
species $a$ and $N_a$ is the number of particles of species $a$. 
The ratio fluctuations $\langle \delta R^2\rangle$ are defined by
\be
\langle \delta R^2\rangle = \langle R^2\rangle - \langle R\rangle^2\quad,
\ee
where the $\langle \dots \rangle$ denotes event averaging.
The ratio $R = N^+/N^-$ is obtained for each event separately 
in the given rapidity interval $\Delta y$ by 
counting  the charges of the hadrons. 
Note that weak decays are not taken
into account - in line with the experimental setups.

As already mentioned  at the beginning of this letter, the predictions for
$D$ obtained in \cite{jeon-koch} are based on
a grand canonical ensemble assuming zero net charge in the system. In 
reality, the system under consideration has a small but 
finite net charge. Also, the assumption of a grand canonical ensemble is only
justified in the limit of   
$P \rightarrow 0$.

For a finite acceptance, transport model calculations and
the experimental data need to be
corrected in order to be compared with the predictions
of \cite{jeon-koch} (see Eq. \ref{eq:correct}).
One observation is that as $P\to 1$, the charge
fluctuation has to vanish because the global charge does not fluctuate.
Also Eq.(\ref{eq:dR2eqdQ2}) is no longer valid in the same limit
if the net charge of the system under consideration is non-zero.

The complete analysis is fully explained in the upcoming 
paper \cite{upcoming}.
In this letter, we list the key results.  
In terms of $N_+$ and $N_-$, the ratio fluctuation is given by
\cite{jeon-koch0}
\be
\ave{\delta R^2}_{\Delta y}
=
\tilde{R}_{\Delta y}^2\,
\left\langle
\left(
{\delta N_+\over \ave{N_+}_{\Delta y}}
-
{\delta N_-\over \ave{N_-}_{\Delta y}}
\right)^2
\right\rangle_{\rm \Delta y}
\label{eq:DefR2}
\ee
where $\ave{\dots}_{\Delta y}$ is the event average in a given rapidity
window $\Delta y$ and
\be
\tilde{R}_{\Delta y} \equiv \ave{N_+}_{\Delta y}/\ave{N_-}_{\Delta y}
\;.
\ee
In an ideal detector
with full coverage, the ratio fluctuation is 
 \be
 \ave{\delta R^2}_{\rm total}
 & = &
 w_H \, \tilde{R}_{\rm total}^2\, 
 {Z_0^2\over\ave{N_-}_{\rm total}\,\ave{N_{+}}_{\rm total}^2}
 \label{eq:D_+-2}
 \ee
 where $Z_0$ is the fixed total charge in the system, 
 $w_H = \ave{\delta N_{-}^2}/\ave{N_{-}} \approx 1$,
 Here $\ave{\dots}_{\rm total}$ denotes the event average in the full
 phase space.
 At SPS, $\ave{\delta R^2}_{\rm total} \sim 1\%$ 
 and at RHIC, 
 $\ave{\delta R^2}_{\rm total} \sim 10^{-5}$.
 This is a consequence of the charge conservation. No
 detailed information on dynamics can be obtained from Eq.(\ref{eq:D_+-2}). 

Having a finite rapidity window modifies the above full phase space result in
an essential way.  
 The particle number in a given rapidity bin now
 has two independent source of fluctuations:
 \begin{itemize}
 \item[(i)]
 Event-by-event fluctuation in the overall number $N_r$
 of a given particle species $r$ due to different impact parameters and
 inelasticities.   
 \item[(ii)]
 Given a class of events with the same $N_r$, 
 additional fluctuations in a specific rapidity window occur
 due to different distributions of the $N_r$ particles in phase space.
 \end{itemize}
 These two fluctuations are independent of each other.
 The squares of the fluctuations are additive.
 Note that the grand canonical ensemble approach 
 requires $\Delta y \ll y_{\rm total}$, therefore the 
 fluctuations from (i) are negligible as shown below.
 
 Assuming that the fluctuations (ii) follow a binomial distribution
 and that the probability $P$ is the same for all species of particles,
 the number fluctuations in a given rapidity window are 
\bea
\ave{\delta N^2}_{\Delta y}
& = &
P^2 \ave{\delta N^2}_{\rm total} + P(1-P)\ave{N}_{\rm total}
\nonumber\\
& = &
P^2 \ave{\delta N^2}_{\rm total} + (1-P)\ave{N}_{\Delta y}
\;,
\label{eq:dy_fluct}
\eea
where $\ave{\dots}_{\Delta y}$ denotes the event average within $\Delta y$
and we used $\ave{N_{\rm ch}}_{\Delta y} = P\ave{N_{\rm ch}}_{\rm total}$.

In the ratio $\ave{\delta N^2}_{\Delta y}/\ave{N_{\pm}}_{\Delta y}^2$ 
needed to construct $\ave{\delta R^2}_{\Delta y}$ (c.f. Eq.(\ref{eq:DefR2})), 
the $P^2$ factor in the first term cancels.  
(The same holds for the cross terms.)
Hence, the contribution of the first term is
$\ave{\delta R^2}_{\rm total}$ in Eq.(\ref{eq:D_+-2}).
The second term replaces $\ave{R}$ factor in Eq.(10) 
in Ref.\cite{jeon-koch0} to yield
 \be
 \ave{\delta R^2}_{\Delta y}
 & = & 
 \ave{\delta R^2}_{\rm total}
 \left(
 { \tilde{R}_{\Delta y}^2 \over \tilde{R}_{\rm total}^2}
 \right)
 +
 (1 - P)\tilde{R}_{\Delta y}^2
 D_{+-}^2\; ,
 \label{eq:Dpm_partial}
 \ee
 where $D_{+-}^2$ is the expression given in Eq.(13)
 in Ref.\cite{jeon-koch0}.
 
 The prediction given in Ref.\cite{jeon-koch} is for 
 $\ave{N_{\rm ch}}_{\Delta y}D_{+-}^2$.
 To extract $D_{+-}^2$ from Eq.(\ref{eq:Dpm_partial}), 
 we need to apply two correction factors, 
 $C_\mu = \tilde{R}_{\Delta y}^2$ and $C_y = (1-P)$.
 Thus, collecting the correction factors yields 
 \be
 {\ave{N_{\rm ch}}_{\Delta y}
 \ave{\delta R^2}_{\Delta y}\over C_\mu C_y}
 = 
 \ave{N_{\rm ch}}_{\Delta y}D_{+-}^2
 +
 { P\over (1 - P)}
 \left(
 \ave{N_{\rm ch}}_{\rm total} 
 {\ave{\delta R^2}_{\rm total} \over \tilde{R}_{\rm total}^2}
 \right)
 \approx
 \ave{N_{\rm ch}}_{\Delta y}D_{+-}^2\; ,
 \ee
 where $\ave{N_{\rm ch}}_{\Delta y} = P\ave{N_{\rm ch}}_{\rm total}$ is used.
 For $P < 1$, the second term is negligible at SPS and at RHIC.

Let us summarize the correction factors that need to be applied 
to take care of the effects of the finite net charge and the finite
acceptance window.
\begin{enumerate}
\item
In order to correct for the finite net charge within the acceptance due to  
baryon stopping, one has to apply a factor $C_\mu$ given by
\bea
C_{\mu} 
=
\tilde{R}^2_{\Delta y}=
{\ave{N_+}^2_{\Delta y} \over \ave{N_-}^2_{\Delta y}}
\label{eq1}
\eea
to the experimental data and the model calculations to compare with the
pion gas and quark gas result of \cite{jeon-koch}. 
\item
In order to correct for the finite bin size in rapidity, 
and in order to incorporate global charge conservation
one has to rescale the experimental data and the transport model predictions by
a factor of 
\be
C_y = 1 - P
=
1-\frac{\langle N_{\rm ch}\rangle_{\Delta y}}
{\langle N_{\rm ch}\rangle_{\rm total}}\quad.
\label{eq2}
\ee
\end{enumerate}
Thus, the basic observable to be compared with the predictions 
calculated in \cite{jeon-koch} is
 \be
 \tilde{D}
  = 
 {\ave{N_{\rm ch}}_{\Delta y} \ave{\delta R^2}_{\Delta y}\over C_\mu\, C_y} 
 =
\left\{
\begin{array}{ll}
1 & \hbox{quark gluon gas}\\
2.8 & \hbox{resonance gas}\\
4 & \hbox{uncorrelated pion gas}
\end{array}
\right.
\label{eq:correct}
 \ee
Again the subscript $\Delta y$ denotes the average taken in the 
rapidity acceptance, while the subscript `total' indicates the 
average of $4\pi$ acceptance.

Fig. \ref{fluc} shows the $\tilde{D}$ values as predicted by the 
UrQMD model (symbols)  as 
compared to the estimates for the resonance gas (dashed line) and the quark
gluon gas (full line). The shown acceptance cuts are chosen according to the
experimentally accessible rapidity windows of the NA49 ($2.5\leq y\leq 4.5$) 
and STAR ($-1\leq y\leq 1$) detectors.
One observes, that the UrQMD model predictions of the charge ratio 
fluctuations are
in agreement  with the expectations of a hadron gas. The predicted
fluctuations are a factor of
3 larger than the fluctuations expected if a QGP was formed.
Thus, the proposed observable -- the charge ratio fluctuations --
predicted for a QGP can not be mimicked by either multiple string break-up nor
strong rescattering effects which may drive the system towards an 
equilibrated Hagedorn gas behavior. 

Fig. \ref{flucdy} explores the fluctuation parameter 
$\tilde{D}$
as a function of the width of the inspected rapidity window 
($y_{\rm cm}\pm \frac{\Delta y}{2}$) in Au+Au, $b\leq 2$~fm 
at $\sqrt s=200$~AGeV. 
Full squares denote the charge ratio 
fluctuation values obtained with all corrections
included as
discussed above. Open squares show the charge ratio 
fluctuation values without the
correction for finite rapidity window and net charge.
In line with the findings of \cite{kk}, one observes a strong decrease in the
fluctuation values if no correction is applied.\footnote{
In Ref.\cite{kk} the authors used PYTHIA to calculate 
$\ave{N_{\rm ch}}_{\Delta y}\ave{\delta R^2}_{\Delta y}$.  Here we are
using UrQMD.  Nonetheless, we get identical results as shown in 
Fig.~\ref{flucdy}.
}
However, the inclusion of the necessary corrections (see
Eqs. \ref{eq1} and \ref{eq2}) yields fluctuations similar to the ones obtained
from a resonance gas over all inspected rapidities.
Notice further the increase of the values obtained from UrQMD
in rapidity windows of $\Delta y \le  2$. 
This increase is due to the vanishing correlations from 
resonance decays which is present at larger rapidity windows.
Details are discussed in \cite{jeon-koch0} (see also
\cite{heiselberg-jackson}). 

In conclusion, the charge ratio fluctuations at SPS and RHIC energies 
are predicted from the UrQMD model and compared to a hadron gas and QGP
estimate.  It is shown that the UrQMD model results are
compatible with the formation of
a hadron gas at SPS and RHIC energies.
The transport model simulations predict fluctuations that are by
a factor of 3 larger then the fluctuations 
characteristic for QGP formation.
The dependence of the fluctuation on the rapidity width is predicted and shown
to be compatible with a resonance gas if all corrections are included. 
This observable can be easily 
studied as `Year 1' observable with STAR at RHIC. 
It can also be directly accessed by the NA49 experiment at the CERN-SPS.

Finally let us stress that it is the corrected observable $\tilde{D}$
defined in Eq. (\ref{eq:correct}) that must be compared 
with our predictions. 
A measurement of $\tilde{D} \simeq 1$ indicates the presence of a
QGP state in the system created by the heavy ion collision.

\section*{Acknowledgements}
This work was supported by the Director, 
Office of Science, Office of High Energy and Nuclear Physics, 
Division of Nuclear Physics, and by the Office of Basic Energy
Sciences, Division of Nuclear Sciences, of the U.S. Department of Energy 
under Contract No. DE-AC03-76SF00098.
M.B. was further supported by the A. v. Humboldt foundation.
This research used resources of the
National Energy Research Scientific Computing Center (NERSC).

\newpage
\begin{figure}[t]
\vskip 0mm
\centerline{\psfig{figure=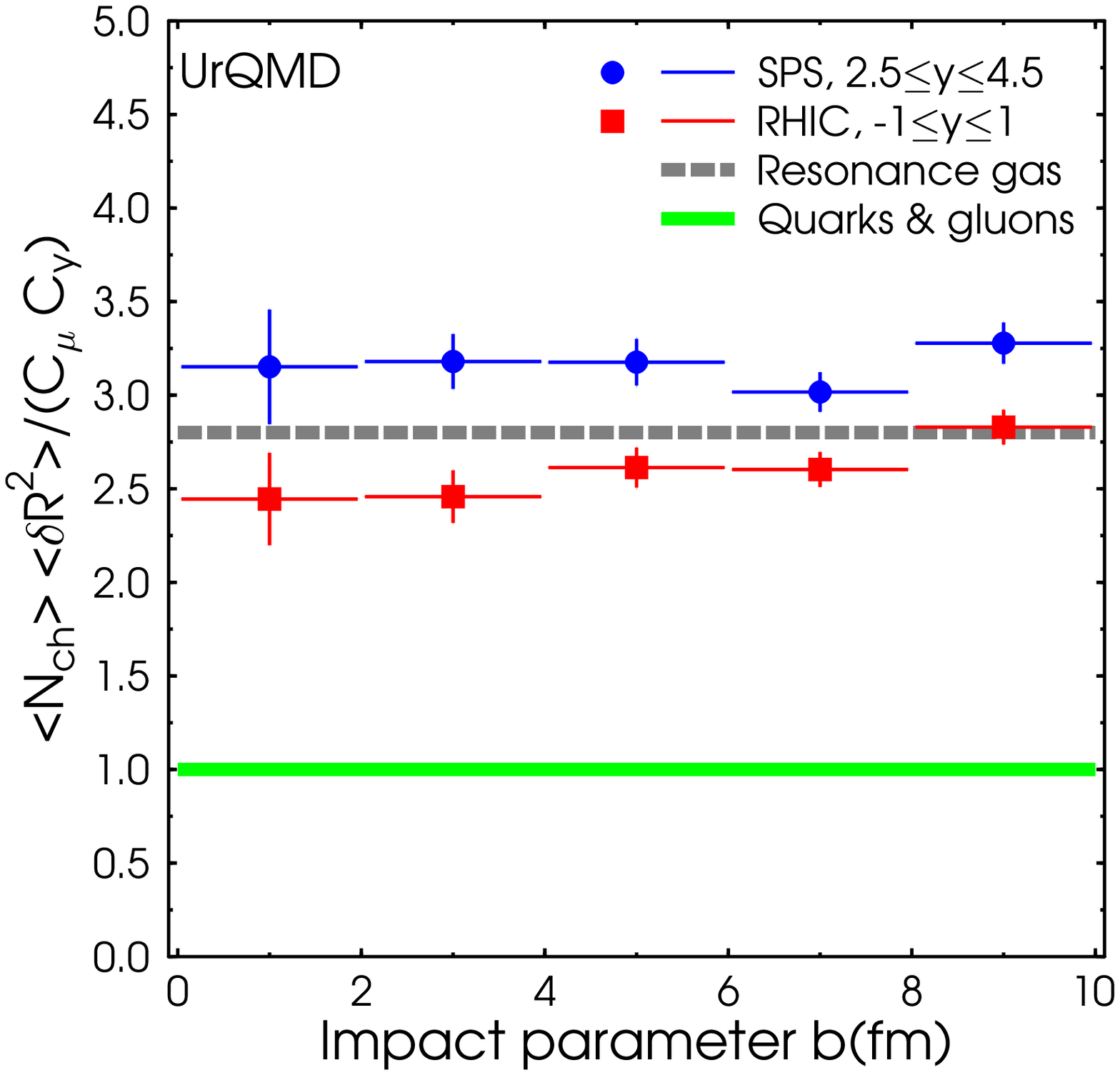,width=15cm}}
\vskip 2mm
\caption{The corrected fluctuation parameter 
$\tilde{D} = \langle N_{\rm ch}\rangle_{\Delta y}
\langle \delta R^2\rangle_{\Delta y}/(C_{\mu} C_y)$ 
as a function of centrality. Circles denote Pb+Pb collisions 
at 160AGeV, while squares denote Au+Au collision 
at $\sqrt s=200$~AGeV as predicted by the UrQMD model. 
The dashed line show the  analytical predictions for a 
hadron gas \protect\cite{jeon-koch},
while the full line depicts the value 
obtained from lattice QCD \protect\cite{jeon-koch}. 
\label{fluc}}
\end{figure}

\newpage
\begin{figure}[t]
\vskip 0mm
\centerline{\psfig{figure=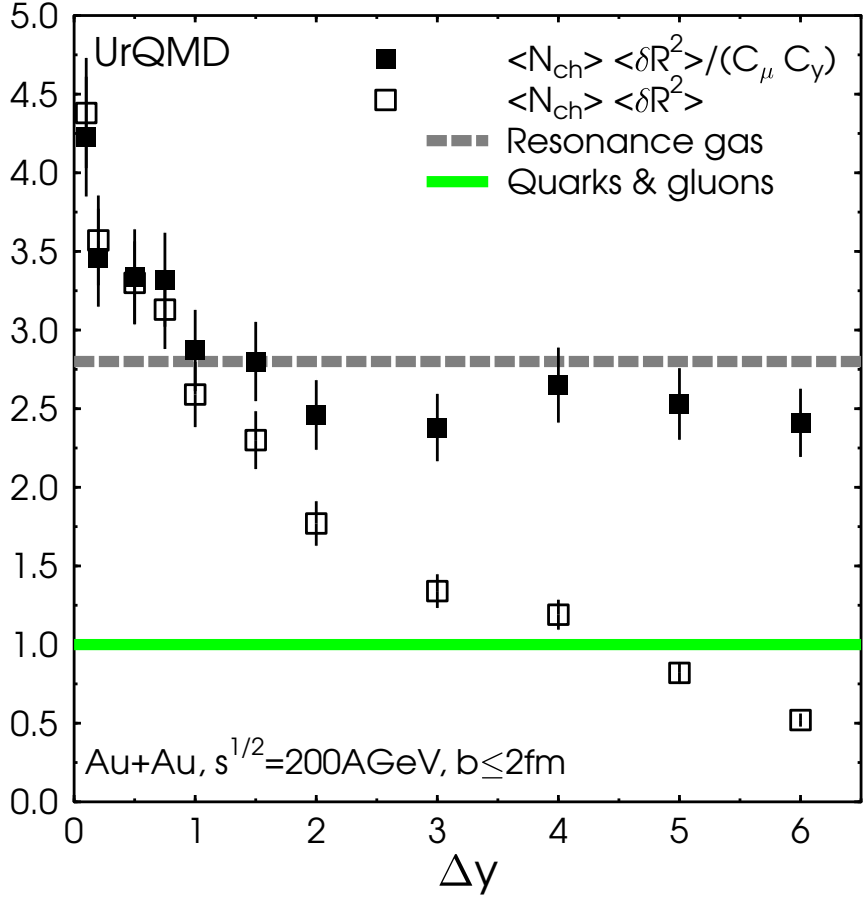,width=15cm}}
\vskip 2mm
\caption{Fluctuation parameter 
$\tilde{D} = \langle N_{\rm ch}\rangle_{\Delta y}
\langle \delta R^2\rangle_{\Delta y}/(C_{\mu} C_y)$ and
$D = \langle N_{\rm ch}\rangle_{\Delta y}
\langle \delta R^2\rangle_{\Delta y}$ 
as a function of the width of the rapidity window 
($y_{\rm cm}\pm \frac{\Delta y}{2}$) in Au+Au, $b\leq 2$~fm 
at $\sqrt s=200$~AGeV as predicted by the UrQMD model. 
Full squares denote the charge fluctuation values $\tilde{D}$
obtained with all corrections included.
Open squares show the charge fluctuation values $D$ without the
correction for finite rapidity window and net charge.
The lines are the same as in Fig. \ref{fluc}.
\label{flucdy}}
\end{figure}

\end{document}